\documentclass[12pt,preprint]{aastex}
\usepackage{amssymb,epsfig}

\newcommand{\rem}[1]{ }
\newcommand{\beq}{\begin{equation}}
\newcommand{\eeq}{\end{equation}}
\newcommand{\bea}{\begin{eqnarray}}
\newcommand{\eea}{\end{eqnarray}}
\def\cite{\citep}

\begin{document} 

\title{Do extragalactic cosmic rays induce cycles in fossil diversity?} 

\author{Mikhail V. Medvedev and  Adrian L. Melott}
\affil{Department of Physics and Astronomy, 
University of Kansas, KS 66045}



\begin{abstract}
Recent work has revealed a
62$\pm$3-million-year cycle in the fossil diversity 
in the past 542 My, however no plausible 
mechanism has been found. 
We propose that the cycle may be caused by 
modulation of cosmic ray (CR) flux by the Solar system 
vertical oscillation (64 My period) in the galaxy, 
the galactic north-south anisotropy of CR production in the 
galactic halo/wind/termination shock (due to the galactic 
motion toward the Virgo cluster), and the shielding 
by galactic magnetic fields. 
We revisit the mechanism of CR propagation 
and show that CR flux can vary by a factor of about 4.6 and reach a
maximum at north-most displacement of the Sun.
The very high statistical significance of (i) the phase agreement 
between Solar north-ward excursions and 
the diversity minima and (ii) the correlation of the magnitude 
of diversity drops with CR amplitudes through 
all cycles provide solid support for our model.
Various observational predictions which can be used to confirm or 
falsify our hypothesis are presented.
\end{abstract}
\keywords{astrobiology --- Galaxy: general --- shock waves --- acceleration of particles --- diffusion --- magnetic fields }

\section{Introduction}

\citet{1} (hereafter RM) performed Fourier analysis of 
detrended data from Sepkoski's compendium \cite{42} 
and found a very strong peak at a period of about 62~My. 
Monte Carlo simulations based on random walk models with permuted 
steps reveal a 99\% probability that any such major spectral peak 
would not arise by chance, thus putting the diversity  
cyclicity on a firm statistical basis. The signal found by RM is
robust against changes in procedure to improve aspects of uneven sampling
and alternate methods of assessing significance \cite{Cornette07, Lieberman+07}.
More recently \citet{Ding} 
have identified a 61 My
periodicity of lower significance in animal transmembrane gene 
duplication events, offset
in phase about 2~radians from biodiversity; the peak in gene duplication events 
corresponds with the 
onset of the declining phase of biodiversity.
Also, \citet{rohde} has noted the same sort of periodicity in
$^{87}Sr$/$^{86}Sr$ isotopic ratios, which is generally taken
as a proxy for continental weathering rates. 
\citet{40} suggested a lack of significant
long-term memory in the most of the fossil record, but they did not
detrend nor
examine lags beyond about 25~My. RM also argued 
that the five great mass-extinctions \cite{43} may be an aspect of 
this cycle. It is very interesting that the 62~My timescale is close 
to the current best value (63.6~My) for the period of the oscillation of the 
Sun in $z$, the distance perpendicular to the galactic disk \cite{2}.  
The Sun is currently about 10~pc north of the plane, moving
away, in an oscillation with amplitude about 70~pc. Finding
a plausible mechanism tied to this vertical oscillation has been
problematic. The primary reason is that midplane crossing (a possible
time of enhanced interactions with galactic matter) would occur 
approximately every 
32~My, which is not a spectral feature RM noted 
as significant in the diversity data.
The same 32~My periodicity occurs if biological
effects are strongest farthest from the galactic plane. 
A recently noted correlation between genus-level diversity and the 
amount of marine sedimentary rock outcropping has been taken as 
evidence that sampling bias may have led to the 
signal discussed here \cite{38}. However, the extent and the lag between the timings
genus diversity and rock outcrop curves and other factors
suggest a common cause for these processes \cite{41,Peters06}. 
Even if measured fossil biodiversity fluctuations are a 
consequence of the claimed periodic sampling bias,
then a strong 62 My periodicity in rock outcropping would 
equally demand explanation.
In general, the idea of astronomical forcings has been given 
a recent boost by a strong result
showing rodent biodiversity fluctuations coupled to 
Milankovic cycles \cite{vanDam}, with a similar periodicity noted
for many clades elsewhere \cite{Martin}.  In what follows, we propose a possible
explanation for this periodicity: enhanced cosmic rays when the solar system
moves to the north side of the galactic plane.

\section{Biodiversity and Cosmic Rays}

Cosmic rays (CRs) may have many different strong biological and climatic effects.
We do not advocate any {\it particular} mechanism for
CR influence on the Earth biosphere. There are many
such mechanisms which can impact fossil diversity, impacting species
extinction and origination.

The first is direct radiation:
CRs produce avalanches of secondary energetic 
particles \cite{13}, which are dangerous or lethal to some organisms.  
If the energy of the primary is below $10^{14}$~eV, only energetic muons
can reach the Earth's surface (some of the muons decays into electrons
and positrons). Primaries with higher energies are able to produce air 
showers that reach the sea level and deliver energetic nucleons as well.
(However, isotopes created by spallation typically have lifetimes 
of order 1 My or less, so
that long-term oscillations in flux would be very difficult to detect.)
Overall, secondary muons are responsible for about 85\% of the total 
equivalent dose delivered by CRs. CR products account for $30-40\%$ of the
annual dose from natural radiation in the US \cite{25}. 
There is almost no protection from muons because of their very high 
penetrating depth, $\sim2.5$~km in water or $\sim900$~m in rock.
Muons are known to produce mutations \cite{micke}.
CRs are therefore a source of DNA damage causing mutations, cancer, etc.
even for deep-sea and deep-earth organisms \citep{Karam+01}. It is beyond the scope
of this paper to compute the absolute increase in the mutation rate. There are many
causes of mutations.  Generally, chemical mutagens induce point mutations, whereas ionizing radiation gives rise to large chromosomal abnormalities. Point mutations usually affect the
operation of a single gene, but large-scale changes in chromosome structure can affect the functioning of numerous genes, resulting in major phenotypic consequences \cite{lodish}.

The second mechanism is climate change:  There is good evidence that
the ions produced by CRs in the atmosphere increase
clouds \cite{22,23,Svensmark+05,Harr+05,6,37} which will increase planetary 
albedo.  It has been argued that this can cool the climate. The global climate response should include changes in temperature and/or precipitation, but the amplitude is highly
uncertain.  

The effect of temperature on biodiversity is emprirically determined.  The tropics
have higher biodiversity, and recovery from mass extinctions is usually taken
by repopulation from the tropics \cite{jab06}.  Furthermore, time series analysis
of green algae, a particularly well-documented clade, shows correlation of 
biodiversity with global temperature over the last 350 My \cite{aguirre05}.  Cool climates are usually drier, since evaporation from the ocean is reduced, and drought typically reduces biodiversity as well. So, the sign of the probable effect of increased cloud formation by CR is in agreement with the sign of biodiversity fluctuations.  This is an area of active research; for example a new experiment startup at CERN (CLOUD) is designed to study the microphysics of cloud seeding by cosmic rays, and much more should be known in a few years. 

The third mechanism is the mutagenic effect of oxides of nitrogen, NO$_x$ (e.g. \cite{Coon}.
CR ionization triggers lightning discharges \cite{24}, which in turn
affect the atmospheric chemistry (e.g., the ozone production by 
lightning and destruction by lightning-produced NO$_x$).
CRs also increase the production of NO and NO$_2$ by direct
ionization of molecules.  Modifications in atmospheric chemistry by
ionizing radiation, and the precipitation of NO$_x$ 
as nitric acid can be numerically modeled \cite {Melott+05, Thom07}.  We plan
to model this in detail based on the CR spectrum expected in our hypothesis.

The fourth mechanism is the mutagenic and damaging effect of solar UVB when
the aforementioned NO$_x$ damage the ozone shield.
Both the indirect and direct production 
of these compounds catalyzes the depletion of ozone, making the 
atmosphere much more transparent to UVB (290-320~nm), 
which can cause mutations, cancer, and kill the phytoplankton
which are at the the base of most of the marine food chain \cite{Melott+05}.
Terrestrial effects of variable CR flux
have been discussed in the context of supernova explosions \cite{17,geh}
and the Sun's motion in the local interstellar medium \cite{29}. 
These models produce random variations on time-scales of hundreds of thousands
of years, so they cannot explain a much longer periodic signal.
The idea of CR-diversity connection lies in line with the newly 
developed approach to mass extinctions as being produced by a combined 
effect of impulsive events and long-term stress called a ``press'' 
\citep{ArensWest06}. Modulated flux of CR provides just such a long-term
and periodic press. 

One interesting possibility is the existince of multiple steady states when there exists
a stronger continuous source of NO$_x$ than exists in the present terrestrial atmosphere.
There is a second steady state with an NO$_x$ density several orders of magnitude higher
than exists at present, which would have catastrophic effects not yet seen in simulations of
astrophysical ionizations to date \cite{kast}.  It will require substantial atmospheric ionization in the troposphere, probably coupled with an additional impulsive event to push the atmosphere toward the 
second attractor.

To summarize, a strong increase in CR flux may affect biodiversity by (1) 
direct radiation effects (mostly by muons)
on the ground and in the seas down to perhaps 1 km, by (2) substantial climate change induced by
cloud seeding of ionization, by (3) the chemical effects of atmospheric NO$_x$ and its rainout as
nitric acid, and by (4) increased solar UVB resulting from ozone depletion, a known effect of
ionizing radiation on the atmosphere.  It will require detailed research to quantify each of
these effects.

\section{Galactic Shock Model}

Low-energy CRs with $10^{10}\textrm{ eV}\lesssim E \lesssim 10^{15}$~eV 
(below the ``knee'') are thought to be produced by galactic sources: 
supernova explosions, supernova remnant shocks, pulsars \cite{17,18}
(hence, referred to as galactic CRs),
whereas higher-energy CR flux is dominated by particles accelerated
in the galactic halo \cite{3} by the shocks in the galactic 
wind \cite{14,30,31} and at the termination shock \cite{18}.
(The boundary of $E\sim 10^{15}$~eV is imprecise: the galactic component
likely extends to $\sim10^{16}$~eV or even higher.) 
The galactic termination shock occurs
when the fast, supersonic galactic wind interacts with the ambient
intergalactic medium, much like the Solar wind termination 
shock on the outskirts of our Solar system \cite{26}.
The position of the shock, which strongly depends on the 
properties of the ``warm-hot intergalactic medium'' \cite{9,8,20} (WHIM) 
and the wind speed, has been estimated \cite{18} to be $R\sim100-200$~kpc 
for the wind speed $V\sim300-500$~km/s. For these parameters with 
the Bohm diffusion coefficient, the extragalactic CR (EGCR) 
flux \cite{jokipii} with $E<E_c\sim10^{15}$~eV was expected to be attenuated
by strong outward advection \cite{31}. However, the first
measurement \cite{19} of the wind speed yielded a 
much smaller value, $\sim100$~km/s (less than the escape 
velocity from the galaxy, hence the outflow is called the 
``galactic fountain''). This puts the shock a factor 
of ten closer, hence decreasing the advection cutoff energy, $E_c$, 
by a factor of 30. Moreover, using a more realistic dependence of 
the diffusion coefficient on particle's energy, $D\propto E^s$ with
$s\simeq0.3-0.6$ (for Bohm diffusion, $s=1$), yields the overall decrease
of $E_c$ by a factor of $10^3$ to $10^5$. Thus, the galactic termination 
shock should be a natural source of EGCRs with energies as low
as $\sim10^{12}$~eV, i.e., those which produce muon
showers in the Earth atmosphere. This EG component is, likely, subdominant 
at the present location of the Sun because of efficient shielding
by galactic magnetic fields, but can be strong at large distances
from the galactic plane, as we will show below. 
CR with energies around $10^{12}$~eV are the most dangerous 
to the Earth biota because they and their secondaries
have the largest flux in the lower atmosphere: lower-energy ones
are attenuated by the Earth magnetosphere, whereas the flux of the
higher-energy particles rapidly decreases with energy.

\begin{figure}
\epsfig{file=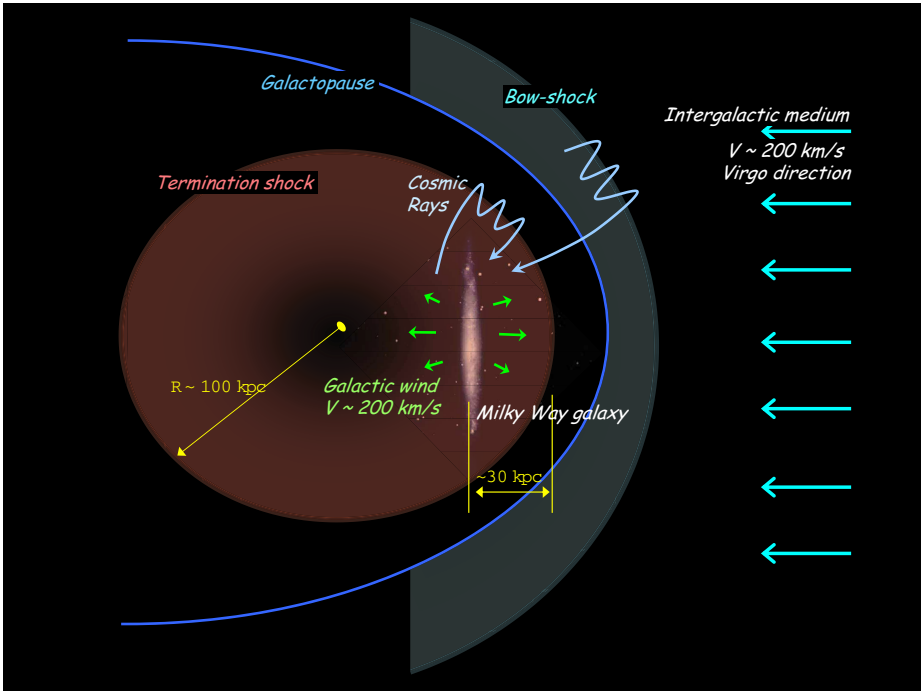, width=122mm}
\caption{Cartoon representing the ``galactosphere'' with the galactic 
termination and bow shocks being sources of extragalactic cosmic rays.
Due to inherent asymmetry, the north side of the Milky Way (with
Virgo cluster being nearly at the north galactic pole) is exposed to 
a larger cosmic ray flux than its south side.}
\label{fig:0}
\end{figure}

The global geometry of the ``galactosphere'' (by analogy with the heliosphere)
is illustrated in Figure \ref{fig:0}. The geometry of the termination and
bow shocks causes the anisotropy
of EGCRs around the Milky Way. In turn, the interaction of the 
gaseous envelope of the galaxy with the WHIM determines the shock geometry.
The WHIM was formed by shock-heating in the early stages of cosmological 
structure formation and should pervade the connected large-scale structure
predicted to form \cite{10} in the Cold Dark Matter
scenario.  Our galaxy moves at the
speed of $\sim150-200$~km/s toward the Virgo Cluster \cite{20,21}, 
which is close to the galactic north pole.
 
The local WHIM is substantially pressure supported,
thus having smaller infall velocity. Motion of the 
galaxy through the WHIM, at even moderate relative velocity, 
pushes the termination shock close to the north galactic face.
Since the center of mass of the Local Group is at low galactic latitude,
our galaxy should not be significantly shielded from interaction with with
the WHIM pervading the Local Supercluster.
The more moderate motion of the Solar
system through the local interstellar medium, $\sim23$~km/s
(c.f., the Solar wind speed is $\sim700$~km/s), produces 
strong asymmetry, with the shock distance in the ``nose'' and  
``tail'' directions differing by more than a factor of two \cite{26,027}.
Therefore, the EGCR flux incident on the northern galactic hemisphere
must be substantially larger than
on the southern hemisphere. The predicted 
strong anisotropy CRs with $E\sim10^{12}-10^{15}$~eV 
is outside the galaxy. At present
Sun's location --- near the galactic plane --- the magnetic shielding
is very strong (as is discussed below), therefore the observed
anisotropy should be very small and, likely, dominated by the 
(nearby) galactic sources. For higher-energy particles of energies about 
$10^{15}$~eV and above, 
our model agrees with previous studies, which predict that 
the CRs are not effectively trapped in the galactic wind;
therefore the anisotropy is intrinsically small.  

In order to see substantial periodic variation in the fossil record, 
the CR flux should have strong 
variation as well. We now demonstrate that the shielding effect 
provided by the galactic magnetic fields against EGCRs
produces the required variation.

CRs with energies below the knee will propagate diffusively through the
galaxy, e.g., in the vertical direction: from the north 
galactic face to the south, which results in (partial) shielding.
A naive application of the standard diffusion 
approximation yields linear variation of the CR density as a
function of $z$. Then the maximum variation of the EGCR flux on Earth 
would be $\sim\Delta/H\sim5\%$, -- too small to have strong
impact on climate and biosphere [where $\Delta\simeq70$~pc is the
amplitude of the Sun vertical oscillation and $H\sim1.5$~kpc
is the exponential scale-height of the galactic disk region 
dominated by magnetic fields \cite{39}]. This picture misses the 
fact that the magnetic field fluctuations in the galaxy are of 
high-amplitude \cite{39}, with $\delta B/\langle B\rangle\sim$few, 
and are likely Alfv\'enic in nature. Therefore, the effects of 
particle trapping and mirroring \cite{35,36} are important.

\begin{figure}
\epsfig{file=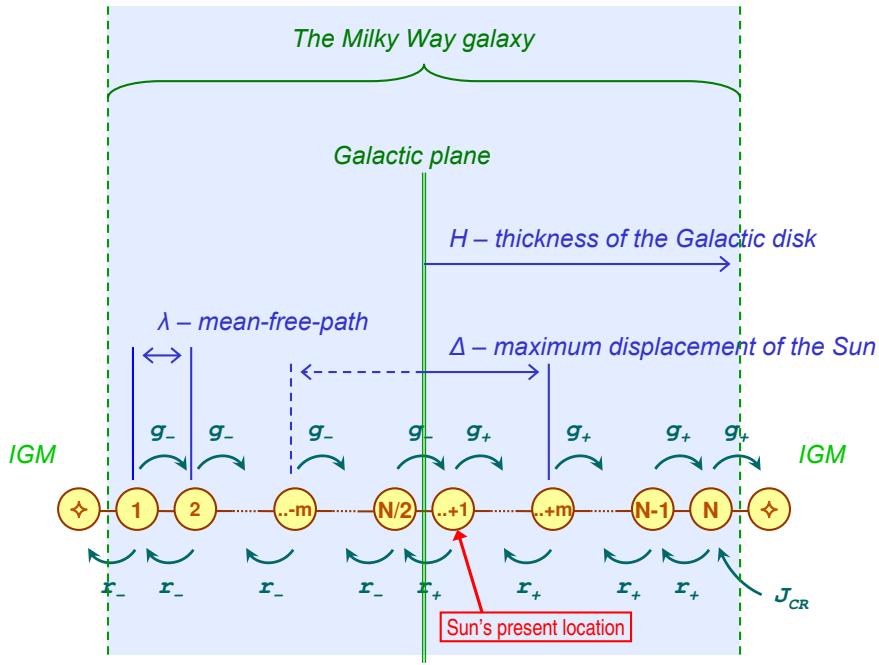, width=122mm}
\caption{ The galactic Markov chain.
The cartoon represents the Markov chain model used to calculate
Cosmic Ray diffusion through the Milky Way galaxy. The chain consists
of $N$ normal sites and two absorbing ($*$) sites, which model 
particle escape. The transition probabilities are $r$ and $g$; their 
subscripts denote position: above ($+$) and  below ($-$) the galactic 
plane. The in-flux of CRs is $J_{\rm CR}$. The Sun moves through sites
between $N/2-m$ and $N/2+m$ and is presently located near the $N/2+1$ site.}
\label{fig:1}
\end{figure}

\begin{figure}
\epsfig{file=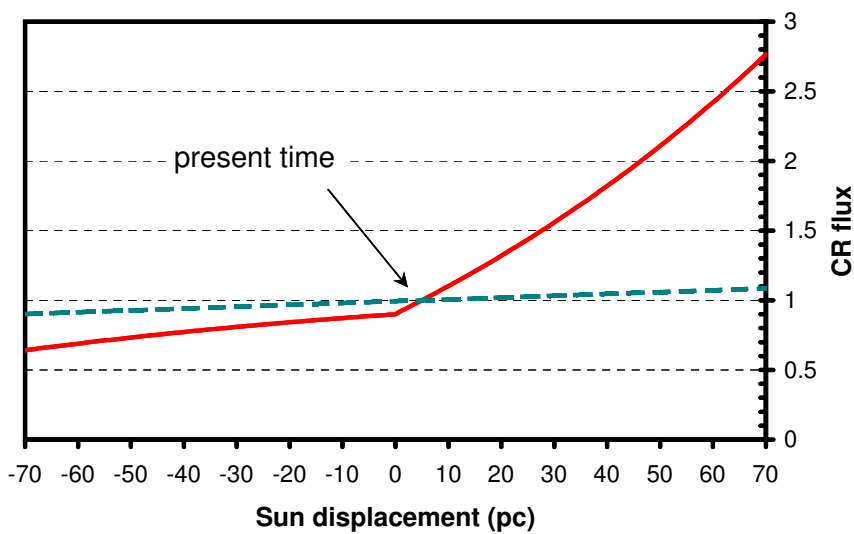, width=122mm}
\caption{The extragalactic cosmic ray flux at the Earth relative to 
the present day value.
The predicted EGCR flux (normalized to the present day value) 
in the Milky Way galaxy as a function of the distance
from the galactic plane (in parsecs) for the asymmetric diffusion
model ({\it solid line}). The standard diffusion model, predicting 
a 5\% increase, is shown for comparison ({\it dashed line}). 
Clearly, a factor of five variation in the EGCR flux at the Earth 
is possible between south-most and north-most excursions of the Sun
from the galactic plane. Here the flux variation, i.e., $F(70)/F(-70)$, 
is about 4.3. Since the Sun's amplitude is a bit larger and varies, the
overall max-over-min CR flux modulation amounts to about 4.6.  }
\label{fig:2}
\end{figure}

We know of no discussion of the effects of transient 
trapping and repeated mirroring in the presence of a mean field gradient 
(as in a galaxy) combined with random walk resulting in {\it asymmetric}
diffusion, in which the probability of particle motion in forward and
backward directions are unequal. This should not be confused with 
the standard diffusion, in which the probabilities are equal, though
the diffusion coefficient can ce inhomogeneous and anisotropic, in general.
The magnitude of the asymmetry is estimated in Appendix \ref{a:transp}.
The number density of CRs in the galaxy is found using the one-dimensional
Markov chain model shown in Figure \ref{fig:1}. 
The galaxy is represented by $N$ sites, separated by one mean-free-path 
distance, thus $N\sim H/\lambda$. The two $*$-states at both ends 
are ``absorbers'' representing escape of CRs from the galaxy. 
The galactic plane is located half-way between $N/2$ and $N/2+1$ sites.
The Sun moves through sites between $N/2-m$ and $N/2+m$, where 
$m\sim\Delta/\lambda$. At present, the Sun is at $z\simeq8$~pc,
which is around site $N/2+1$. The forward and backward transition 
probabilities are $r$ and $g$; their subscripts denote position:
above ($+$) or below ($-$) the plane. 
There is in-flux of CRs, $J_{\rm CR}$, (produced at the
termination shock in the northern hemisphere) through the right end.
The observed CR flux decreases with energy roughly as $\propto E^{-3.1}$ 
above the knee and as $\propto E^{-2.7}$ at lower energies. 
We make a conjecture that the ``true'' EG flux has no break at 
$E\sim 3\times 10^{15}$~eV, whereas the observed break (the knee)
is due to magnetic shielding discussed above. Thus, we conjecture that
the EGCR flux outside the Galaxy exhibits no breaks up to $10^{12}$~eV, 
which could be a local maximum and which could smoothly join the lower
energy calactic component. Thus, the EGCR flux of
the most dangerous particles of $E\sim10^{12}$~eV is about two orders of 
magnitude higher than the observed flux at this energy (though,
it is still too small to affect the galactic structure, see
more discussion in Appendix \ref{a:transp}). We use this value 
for the parameter $J_{\rm CR}$ of our model.
An analytical
solution for the CR density is plotted in Figure \ref{fig:2}.
An exponential increase of the local EGCR density with $z$ is seen. 
For contrast, we also plot the result of the standard diffusion model. 
Thus, very strong exponential shielding from EGCRs is found. 
Our estimates in of the CR flux are somewhat conservative in a number of 
places, so the actual flux may be a factor of few higher, and should depend on
particle energy, the properties of the galactic magnetic fields
and turbulence spectrum. Further details of CR propagation are included in 
Appendix A.

It should be understood that this solution contains only the CR variation due to solar
motion normal to the galactic plane. There is an (unknown) additive component of CR due
to, for example, supernovae. For example, there has apparently been a CR enhancement
the last few My \cite{lavielle}, which is probably due to recent nearby supernovae, for which
there is evidence for at least one about 2.8 Mya \cite{knie}, possibly associated with the 
formation of the Local Bubble in the interstellar medium \cite{jesus,fuchs}.  The long-term 
variation we propose will be superimposed on these shorter-term variations.

\section{Results}

Figure \ref{fig:3} shows the detrended fossil genera fluctuation from 
\citet{1}
and the computed EGCR flux from our model versus time for the last
542 My. The fossil data are timed to within the uncertainties of geological
dating methods. Here we used the best available model data for the
solar position $z$ versus time \cite{2} kindly provided to us by D.~Gies. 
These calculations assume azimuthal symmetry of the Milky Way.
The oscillation period and amplitude varies in response to the 
radial motion of the Sun and a higher density toward the Galactic 
center (included in the calculation) and a
scatter from spiral arm passage (not included), see more discussion
in Appendix \ref{a:motion}.
Note that the long-term modulation of CR maxima in Figure \ref{fig:3} is real,
being due  to the Sun's radial motion. Hence, one should also expect a weaker
long-term cycle with a period $\sim170$~My in the fossil record. 
The average period, accurate to about 7\% \cite{2}, 
$\sim63.6$~My coincides precisely within uncertainty with the 
$62\pm3$~My period of the fossil diversity cycle \cite{1}.
RM noted that the 62~My signal in the fossil 
record emerges from integration over almost 9 periods, and while highly
significant does not coincide exactly with the onset of major extinction 
events, dated to within uncertainties in geological dating methods.
These may be caused by a combination of stresses including for example
CR flux variation,
bolide impacts, volcanism, methane release, anoxia in the oceans,
ionizing radiation bursts from other sources,
etc. (It is an interesting aspect of this
that the onset of the K/T (``dinosaur'') extinction, generally thought to be due to a
bolide impact, coincides within 2~My of mid-plane
crossing \cite{2}.) Nevertheless, the 62~My cycle is strong 
and robust against alternate methods of Fourier decomposition and 
alternate approaches 
to computing its statistical significance \citep{Lieberman+07}.


A number of statistical tests have been performed in order to 
address the significance of the correlation between fossil data and
modelled cosmic ray flux.
In Appendices C and D, we discuss cross-correlation
analyses involving the detrended data, the raw data
for the short-lived genera [both samples are from \citet{1}],
and Fourier-filtered samples. All tests show high statistical
significance of the CR vs. diversity correlation. 
Namely, the detrended sample used by RM
correlates with the CR flux from our model at the level of 49\%
(the Pearson coefficient is $r=-0.49$). 
The diversity data contains 167 discrete time periods; however 
only about 59\% of the fossil data used is resolved to the bin size.  
For a conservative assessment of statistical 
significance, we take the effective number of bins as 
$\sim167\times0.59\approx100$.
The result is very high statistical significance that the observed
correlation is not a 
consequence of coincidence (p-value $1.9\times10^{-7}$).  Consequently
we conclude that the CR flux variation model may explain about one half of
the long-term fluctuations in detrended biodiversity.

\section{Discussion}

The 62~My CR fluctuations are of course not the only source of 
diversity changes, but explain the long-period cycles quite well.
Filtering out the short-term waves (the short-term component is largely 
dominated by the effect of the finite bin size of few My),
the cross-correlation amplitude rises to $\sim57$\% at even higher 
better significance level (the probability of chance coincidence ---
p-value --- is $6\times10^{-10}$), see Appendix \ref{a:corr} for details. 

Our model is predictive. An unavoidable consequence of the model is
that the varying amplitude of solar excursions from the galactic plane
modulates the CR flux and, consequently, affects the magnitude of the 
diversity drop. As we discussed above, variation in CRs are
real in Figure \ref{fig:3}. Thus, we look for possible correlation
of the amplitude of the CR flux in each ``cycle'' and the amount of diversity
drop (which we call ``extinction strength'') in the corresponding 
diversity drop ``event''.  
We applied an algorithm which finds for each CR cycle
the local diversity minimum nearest to the CR maximum and the
nearest preceding diversity maximum. The ``extinction strength'' is
calculated as the difference between these maxima and minima.
The CR amplitude is calculated analogously. 
Details are given in Appendix \ref{a:corr2}.

The first result is that for each CR maximum there is {\it always}
a diversity minimum within few My (the largest ``mismatch'' of 9~My is for 
the K/T extinction, which is likely due to a bolide impact). 
This is a strong result, 
given that some fossil bin sizes are as large as 5-6~My and, moreover, 
that only 59\% of genera are resolved to this level. Second,   
Figure \ref{fig:4} shows an impressively strong correlation of 
the peak CR flux for each cycle in Figure \ref{fig:3} and the 
corresponding diversity drop.
[Some uncertainty in the galactic structure data
(magnetic fields, turbulence, halo structure) can affect the
overall amplitude (normalization) of CR maxima, but not their 
 rank order strengths.]
A similar procedure was also applied to the raw data for short-lived 
genera to avoid biases in deep time due to the cubic fit. 
The results are shown in the Supplementary Information.
The linear correlation coefficient of the CR flux amplitude
and the extinction strengths shown in Figure \ref{fig:4} is 
over 93\% and is significant,
at the level of 99.93\% (p-value of $6.8\times10^{-4}$, computed from the 
Student distribution).
This provides a very solid and independent 
confirmation of the model, which provides a natural mechanism for 
observed cycles in fossil diversity.

It should be noted that the cosmic ray enhancement axis shown in Figure
\ref{fig:4} is a consequence of our assumed normalization 
of the EGCR flux.  While our assumed flux is physically reasonable, 
other values are possible. Changing this assumption would stretch the CR 
axis, but would not change the rank order of the values nor modify their 
agreement with the diversity drop.  Such a change would somewhat affect 
the shape of the CR curves in Figure 3, making a small change in the 
amplitude of the crosscorrelation.  However, the agreement in period, 
phase, and rankorder diversity drops, the major pieces of evidence 
supporting our model, are not affected.

We emphasize that our hypothesis attempts to explain the very long term
variations of the diversity and relate it to the long-term variation of
the CR flux. We, by no means, attempt to explain every possible variation
of the cosmic ray flux, which is known to be affected by galactic
sources. For instance, a spiral arm crossing or a nearby supernova can 
substantially (depending on the distance)  increase the CR flux 
at the Earth for the life-time of 
the supernova remnant, which is tens to hundreds of thousands years. Any such 
variation of CR flux may be ``recorded'', e.g., in Earth isotope abundance,
but has nothing to do with the discussed trend over hundreds of millions 
of years.

Our hypothesis is falsifiable. We predict that the termination
and bow-shocks of our galaxy will be distributed sources of EGCR
up to TeV energies at the galactic north, whereas only EGCR of 
$10^{15}$~eV will be reaching the Galaxy from it's south. 
Such an anisotropy could potentially be observed directly (though
shielding by and trapping in the galactic magnetic fields may 
diminish the effect substantially. A crude estimate yields 
about 1\% anisotrophy, provided the effects of local CR sources
is negligible. There are a number effects produced by CRs, 
some of which are evaluated in Appendix \ref{a:pred}, whereas others
are postponed intil future publications. In particular, we predict
that EGCR will upscatter 2.7~K cosmic microwave background radiation
to the soft X-ray band ($\sim200$~eV). The up-scattered flux shall
constitute about $10^{-7}$ of the observed cosmic X-ray background.
Similarly, EGCR will upscatter far and near infra-red (FIR and NIR) 
photons to about 10~keV and 1~MeV energies, respectively. 
The fraction of the upscattered photons compared to the hard X-ray 
and gamma-ray backgrounds is about $10^{-8}$ in both cases.
Although the relative effects are rather weak, one can look for global 
north-south anisotrophy of cosmic X-ray and gamma-ray backgrounds.
Averaging over half of the sky will substantially lower the statistical 
fluctuations of the backgrounds due to the galactic foreground emission.
Taking the difference of the hemisphere-averaged signals may 
prove to be a useful strategy. Yet another possibility is to look for
pion decay products (gamma-rays and neutrinos) produced in interactions
of CR protons with neutral hydrogen in the Galaxy. We predict
(see Appendix \ref{a:pred}) the emission of $\sim2$~GeV protons with
the flux of $\sim1\%$ of the cosmic gamma-ray background at this energy.
Detection of this excess emission from the northern Galactic hemisphere
seems a feasible task for {\it GLAST}.
Similarly, we predict the excess of $\sim900$~MeV neutrinos with flux
$\sim3\times10^{-4}\ {\rm particles/(m^2\ s\ sr)^{-1}}$, which may 
be a good target for {\it IceCube}.

To summarize, we have suggested that a substantial flux of 
cosmic rays is produced in a shock at galactic north, -- a direction 
toward which our Galaxy has long been known to be moving in the Local 
Supercluster. We have shown that there is considerable shielding 
from the cosmic rays due to the gradient of both the regular and
turbulent magnetic field in the Milky Way.  When this is combined with 
the kinematics of the Sun in the Galaxy, based on its present motion 
and the galactic gravitational field, we find a highly statistically 
significant agreement between period, phase and relative magnitudes of 
excursions of extragalactic cosmic ray intensity and drops in biodiversity 
as measured by marine genera. There are many direct and indirect mechanisms 
by which cosmic rays may affect biodiversity.  
The cosmic ray periodicity discussed here accounts for about one-half of the 
biodiversity variance found by Rohde and Muller, and nearly all of their 
main feature, the 62 My oscillation. The model predicts that higher  
excursions of the Solar system above the galactic plane should produce 
more cosmic ray flux 
and larger biodiversity drops; this prediction is borne out by comparison 
between solar motion and paleontological data.

\acknowledgements
The authors are grateful to R. Muller for providing us with the 
diversity data and to D. Gies for providing the solar motion data.
The authors thank B. Lieberman, B. Anthony-Twarog, J. Ralston, A. Karam, 
B. Fields, J. Scalo, C. Flynn, and R. Bambach for discussions. 
This work has been supported by DoE grant DE-FG02-04ER54790 (to M.V.M.) and 
NASA Astrobiology program grant NNG-04GM41G (to A.L.M. and M.V.M.).

\appendix
\section{Cosmic ray transport through the galaxy}
\label{a:transp}

CRs with energies below the knee in the galaxy propagate
diffusively. The Larmor radii of the particles are smaller than
the field inhomogeneities, so they nearly follow field lines.
These fields are turbulent \cite{32,39}, hence the effective 
diffusion \cite{034,35}. One often assumes the Bohm diffusion coefficient
for this process. As EGCR particles diffuse through
the galaxy (in our case, in the vertical direction, from the north 
face to the south), their density decreases, thus resulting
in shielding. As we pointed out in the text, a naive application 
of the diffusion approximation yields linear variation of the CR density 
with $z$ of about 5\%, for typical galactic parameters \cite{39}). 
High-amplitude magnetic field fluctuations in the galaxy \cite{39}
affect diffusion via mirroring and transient trapping effects \cite{35}.
In the presence of the mean field gradient, they modify diffusion 
so that it becomes asymmetric (not to be confused with anisotropic diffusion,
where diffusion is still symmetric, but the rates depend on position 
and orientation). We are not aware of discussion 
of this effect in the literature.

In asymmetric diffusion, the probabilities of the forward and backward
transitions are not equal.
To estimate the magnitude of the asymmetry, recall that the amplitude 
of turbulent magnetic fluctuations is maximum on large spatial 
scales and decreases as magnetic energy cascades to small scales. 
Hence trapping by large-amplitude waves occurs on scales 
comparable to the field correlation length \cite{32}, hence $\lambda\sim20$~pc.
(The mean-free-path, $\lambda$, depends on particle's energy as well.)
Trapping is intermittent and transient because large-amplitude,
quasi-coherent Alfv\'enic wave-forms (``magnetic traps'' or 
``magnetic bottles'') exist for 
the Alfv\'en time. Thus, a trapped CR particle, moving at almost the 
speed of light, experiences about $N_b\sim c/V_A\sim3\times10^3$ bounces 
(for the interstellar medium field $B\sim3~\mu$G and density 
$\rho\sim3\times10^{-26}$~g/cm$^{3}$, where 
$V_A=B/\sqrt{4\pi\rho}$ is the Alfv\'en speed) during the bottle
lifetime. Reflection conditions are determined by the 
particle loss-cones on both ends of the magnetic bottle. There is a 
field gradient ($B$ decreasing away from the galactic
plane on a distance $\sim H$. The precise value of $H$ 
not known, but it does not significantly affect the results of our model).
The loss-cone conditions
imply that, on average, more particles are reflected from a higher-field 
end (closer to the galactic plane) than from the 
lower-field one. From the loss-cone condition, we estimate the
reflected fraction in one bounce as 
$\epsilon_0\sim\lambda[\langle B\rangle/(\nabla\langle B\rangle)]\sim10^{-2}$.
Since particles also interact with the smaller-amplitude, high-frequency
background of short-scale Alfv\'en waves, the particle distribution
function evolves toward isotropization while trapped particles traverse
the magnetic bottle. We assume some 1\% efficiency of this process,
$\eta\sim0.01$. 
This leads to a small ``leakage'' of particles from
the trap, predominantly in the direction away from the galactic plane.
The total ``leaked out'' fraction per the trap lifetime is
$\epsilon\sim 1-(1-\eta\epsilon_0)^{N_b}$. 
The efficiency depends on the numerous factors, such as
the level of Alfv\'enic small-scale turbulence which induce pitch-angle
scattering of CRs in and out of the loss cones, the relative phase-space
volumes occupied by the loss cones and the trapped regions, the
in- and outgoing fluxes of CRs relative to the local density of 
particles at a given Markov site. The latter depend upon the 
total leaked out faction of particles, thus $\eta$
should be determined self-consistently via numerical modeling.  
A complete 
calculation of all these processes will be presented elsewhere. 

The number density of CRs in the galaxy is found using the one-dimensional
Markov chain model shown in Figure \ref{fig:1} and discussed in the text. 
Note that the forward and and backward transition probabilities 
above and below the galactic plane are $r_+=g_-,\ r_-=g_+$, by symmetry. 
Their ratio is $g_+/r_+\sim1+\epsilon$,
with $\epsilon$ obtained in the previous paragraph. An analytical
solution for the CR density is plotted in Figure \ref{fig:2}.
An exponential increase of the local EGCR density with $z$ is seen. 
There, the result of the standard diffusion model, i.e., with $\epsilon=0$
is also shown. Very strong exponential shielding effect is seen (the EGCR 
flux at low energies is normalized by the present day value to unity). 

The amplitude of the EGCR flux variation is determined by the total
EG flux outside the galaxy and the structure of magnetic fields and
turbulence inside the galaxy. The flux of EGCRs outside the Milky Way
can readily be evaluated, taking into account that the effective 
shielding scale-height is about few hundred pc (a factor of few 
smaller than $H$) because at about 100 pc away from the galactic plane,
the galactic wind is beginning to form (c.f., the thickness of 
the thin galactic disc is $\sim80$~pc). This wind stretches the 
magnetic fields in the $z$-direction (wind direction), thus 
dramatically increasing the field correlation length and 
the particle mean-free-path, $\lambda$, and decreasing the 
turbulence level of small-scale Alfv\'enic fluctuations responsible 
for pitch-angle diffusion. These both effects reduce the exponential 
suppression nearly to the standard diffusion value above $z\sim300$~pc or so.
This extrapolation yields that the EGCR flux outside the Milky Way is
about one hundred times larger than the local value.

The CR flux above the knee (above $10^{15}$~eV) is thought to be primarily 
extragalactic in origin
(these CRs are not trapped in the galaxy because of their large Larmor radii),
and it decreases with energy roughly as $\propto E^{-3.1}$. Below the knee,
the CR spectrum is shallower, namely $\propto E^{-2.7}$, whereas the CR 
particles are ``trapped'' in the galactic magnetic fields. We make a
conjecture that the ``true'' EG flux has no break at 
$E\sim 3\times 10^{15}$~eV, whereas the observed break (the knee)
is due to magnetic shielding discussed above. Thus, the EGCR flux of
the most dangerous particles of $E\sim10^{12}$~eV is about two orders of 
magnitude higher than the observed flux at this energy. This
value of the EGCR flux matches nicely the extrapolated value
of the CR flux discussed in the previous paragraph (as well as with 
the overall energetics of the galactic wind). 

Here we also comment that such high EGCR flux is still too low to 
affect the global galactic structure (via CR pressure).
The CR pressure in the galaxy is dominated by particles 
with energies below tens of MeV per nucleon (the Earth is protected 
from them by the Solar Wind) and constitutes up to ten percent 
of the total pressure. However, we discuss here the much more
energetic particles, with energies above ten TeV,
which are not attenuated by the Solar Wind and the Earth magnetic fields. 
With the galactic CR spectrum $\propto E^{-2.7}$ at low energies,
one obtains that these dangerous TeV EGCRs contribute less than  
0.1\% of the total pressure. This is an upper limit on the EGCR
pressure outside the Milky Way. Inside the galaxy, the EGCR
pressure is lower because of magnetic shielding.
Hence it has no influence on the dynamics of the interstellar medium 
and the galactic structure while having a potentially devastating effect on 
life on the Earth.

The assumed EGCR flux at and above TeV is reasonable on
energetic grounds. Indeed, the kinetic energy density in the 
outflowing galactic wind is of order the energy density of galactic
CRs \cite{18}. Some fraction, $\eta$, of the wind energy goes into 
acceleration of EGCR at the galactic termination shock. 
Thus, one can estimate, by analogy with the previous paragraph, that
the lower limit on the conversion efficiency is about $\eta >10^{-3}$.
This is a very reasonable value, given large uncertainties in the 
galactic halo structure, its magnetic fields, the diffusion
coefficient and its dependence on particle energy, etc. 

Our estimates in of the CR flux are somewhat conservative in a number of 
places, so the actual flux may be a factor of few higher.
We also neglected here that the 
CR flux at Earth depends on the injected energy spectrum at
the termination shock and on the particle mean-free-path
in the galactic fields, which is energy-dependent.
Thus, the amplitude of CR fluctuations should depend on
particle energy, the properties of the galactic magnetic fields
and turbulence spectrum. 

\section{Observational predictions of the model}
\label{a:pred}

The large anisotropy of EGCR could potentially be detected.
Since we are now well inside the galaxy and shielding is very strong,
direct detection of CR north-south anisotropy is complicated.
The expected anisotropy at the present location of the Earth 
can cruedly be estimated from the model parameter $J_{CR}$, the
fraction of CR entering the Galaxy from the north that leave it at the south.
This yelds the anisotropy being of order 1\%. However, this does
not take into account any (local) galaxtic sources of CR that can
``contaminate'' the signal. 
Studies of CR anisotropies indicate their existence at 0.1\%-1\%
level. However, they are mostly attributed to the local magnetic
field structure --- spiral arms. This is reasonable because charged
particles propagate nearly freely along field lines (mostly parallel
to the galactic plane) and diffuse across them in the vertical direction.
Recent {\it Milagro}, {\it Super-Kamiokande} and {\it Tibet Air Shower Array} 
results indicate the presence of the anisotropy 
at the level of 0.1\% in the Galactic center direction and along
the Galactic north-south direction \citep{Atkins+05,Oyama06,Amen06}.
However, the absence of the Compton-Getting effect may imply that 
the galaxy-wide anisotropy could be smeared out by a propagation 
effect through the co-moving local interstellar medium. 

One can think of some indirect methods. For instance, TeV EGCR can
Compton up-scatter CMB photons to energies of 
$
\sim\gamma^2(2.7{\rm K})\sim200~{\rm eV}$. 
The up-scattered flux of photons is of order
$F_{scatt}\sim F_{CMB}\tau$, where the optical depth to scattering is
$\tau\sim R\sigma_{p\gamma}(4\pi F_{CR}/c)$, where in turn
$F_{CR}$ is the EGCR flux outside the Galaxy, $R$ is the typical size 
of the system being of order the distance to the shock, and 
$\sigma_{p\gamma}\sim0.1$~mb is the typical $p-\gamma$ cross-section.
With the conjectured CR flux $F_{CR}\sim10\ ({\rm m^2\ s\ sr})^{-1}$
at 1~TeV and $R\sim30$~kpc, we obtain the upscattered flux of order
$\sim2\times10^{-2}\ ({\rm m^2\ s\ sr})^{-1}$. The
galactic north-south anisotropy of these soft X-ray photons
could be a clear signature of our model. However, detection of
such anisotropy can be difficult because of large contamination
by gas line emission at these energies and strong hydrogen 
absorption of these soft X-rays. Indeed, the cosmic soft X-ray background
(CXB) is about $\sim10^{5}\ ({\rm m^2\ s\ sr})^{-1}$ at 0.5--1~keV
\citep{CXB}. This yields that the upscattered photon flux is
about $10^{-7}$ of the CXB, which makes it hard to observe.

TeV cosmic rays can also up-scatter infrared (IR) photons. 
Cosmic IR background has two peaks, in the far and near infrared,
at approximately $\sim10^{-2}$~eV and 1~eV. Up-scattering of FIR 
and NIR photons brings them to about 10~keV and 1~MeV, respectively.
The FIR and NIR fluxes at the peaks are approximately 
$2\times10^{-8}$ and $3\times10^{-8}$ ${\rm W\ m^{-2}\ s^{-1}\ sr^{-1}}$.
The up-scattered fluxes are calculated to be of order
$4\times10^{-5}$ and $4\times10^{-7}\ {\rm photons\ m^{-2}\ s^{-1}\ sr^{-1}}$.
The ratio of the up-scattered fluxes to the hard X-ray (at 10~keV) 
and gamma-ray (at 1~MeV) background fluxes \citep{CXB,CGB} are
about $10^{-8}$ in both cases. 

Yet another possibility is to 
look for $\gamma$-rays at GeV energies due to interaction of
TeV EGCRs with the interstellar gas in molecular clouds and production 
of pions, which then produce $\gamma$-rays via decay 
($\pi^0$'s are of great interest because their motion is not
affected by the galactic magnetic fields). The $p-p$ cross-section
for neutral (and charged) pions is rougly $\sigma_{pp}\sim30$~mb 
\citep{Dermer86} for 
proton center of mass energies of $\sim30$~GeV, which corresponsd to 
interaction of TeV CRs with neutral hydrogen in the Galaxy. 
For the estimate, we assume the neutral hydrogen column density
$N_H\sim10^{21}$~cm$^{-2}$ (this is somewhat higher than the average
column density toward the Galactic poles). The collisional depth
is $\tau\sim N_H\sigma_{pp}\sim 3\times10^{-5}$. This yields the
gamma-ray flux of about $F_\gamma\sim3\times10^{-4}\ ({\rm m^2\ s\ sr})^{-1}$.
Neutral pions decay $\pi^0\to2\gamma$ to produce $\sim70$~MeV photons
(in the center of mass frame). Thus, we predict emission of
gamma-rays of energy of $\sim2$~GeV from pion decay. The ratio of the
predicted gamma-ray flux to the cosmic gamma-ray background \citep{CGB}
is of order 1\%, which seems feasible to detect. Such an observation 
would be an interesting task for {\it GLAST}. This effect will be 
addressed in a subsequent publication.

Similarly, interaction of TeV CRs with neutral hydrogen produces
charged pions, which decay into muon neutrinos of energy $\sim900$~MeV. 
Similar estimates yield neutrino flux of about 
$F_\nu\sim3\times10^{-4}\ {\rm particles/(m^2\ s\ sr)}^{-1}$ at $0.9$~GeV.
These neutrinos can be observed with {\it IceCube}.

\section{Model of the solar motion through the Milky Way}
\label{a:motion}

The solar motion through the Milky Way has been computed 
for the past 600~My and kindly provided to us by D.~Gies \cite{2}.
A number of axisymmetric galaxy models have been presented and analyzed
by \citet{50}. Gies' computation uses the best  
model of the global density distribution in the galaxy, 
according to the analysis of that paper \cite{50}. The density 
normalization Dehnen \& Binney used is about $0.17~M_\odot~\textrm{pc}^{-3}$,
which is somewhat higher than the local density of 
$0.1\pm0.01~M_\odot~\textrm{pc}^{-3}$ found by these authors \cite{51}
and other groups \cite{52,53,54} from the {\it Hipparcos} 
parallax data. The latter, low value of the galactic density results
in a longer period of vertical oscillations at the present
position of the Sun, as long as $82\pm2$~My, which is substantially
larger than the average period of 64~My. It should be noted, however, 
that {\it Hipparcos} has determined parallaxes and distances to stars 
within 200~pc (in the galactic plane) around the Sun. Even with the very 
low Sun velocity with respect to the local rest frame, $v\sim13$~km/s 
(c.f., nearby stars have typical velocities of about $\sim40$~km/s), 
the Sun will traverse the {\it Hipparcos}-probed region 
within 15~My, much shorter than 64~My average period.
The low local density is consistent with the fact that
the Sun is in the inter-arm region at present. The strength of
the spiral arms is still debated \cite{55}, however recent
Doppler measurement of a maser in the Perseus arm \cite{56}
indicates very strong contrast of the arm---inter-arm density.
With a reasonable 50\% duty cycle (the Sun arm-crossing time vs.
inter-arm residence time) the average oscillation period is
in agreement with 64~My.

\citet{50} used {\it Hipparcos} 
data in their models. However, instead of using the local
galactic density as a normalization, they considered it as a free
(fit) parameter, which they found by fitting observables, e.g.,
the star terminal velocities \cite{57}. These are, in turn, determined from
proper motions found by {\it Hipparcos}. Since such measurements do
not depend on parallax distances, one probes distances as 
large as 3~kpc. Thus, such a technique is much more accurate to
constrain a {\it global} galactic model. Of course, the axisymmetric
model misses the local density inhomogeneities (e.g., due to spiral arms),
which should result in some scatter of the Sun oscillation period.
In fact, the diversity period does show a larger scatter than
the computed vertical oscillation (and the related CR flux).

\section{Statistical analysis}
\label{a:stat}

The correlation between two data sets is evaluated with the
Pearson moment correlation coefficient, $r$, a dimensionless
index that ranges from --1.0 to 1.0 inclusive and reflects the 
extent of a linear relationship between the two sets.
For two sets of $N$ values each, $X$ and $Y$, the 
r-value is calculated as
\beq
r=\frac{N\left(\Sigma\,XY\right)-\left(\Sigma\,X\right)\left(\Sigma\,Y\right)}
{\sqrt{\left[N\left(\Sigma\,X^2\right)-\left(\Sigma\,X\right)^2\right]
\left[N\left(\Sigma\,Y^2\right)-\left(\Sigma\,Y\right)^2\right]
}}.
\eeq 
The statistical significance of the correlation is evaluated from 
the Student t-distribution. The t-distribution is used in the 
hypothesis testing of sample data sets and gives the probability
(p-value) of the chance coincidence. In the limit of large number 
of degrees of freedom (data points), it approaches the Gaussian distribution.
For non-zero $r$,  
\beq
t=\sqrt{{r^2(N-2)}/{(1-r^2)}}
\eeq
obeys the Student's t-statistics with $N-2$ degrees of freedom.

\section{Correlation of CR flux and diversity data}
\label{a:corr}

All the diversity data used in our analysis are taken from
supplementary information files of \citet{1} paper.
The cross-correlation of the predicted CR flux and the de-trended
diversity data is discussed in the text. One can worry that 
the de-trending strongly affects the data and can introduce
biases. In their original paper, RM demonstrated that
the 62~My periodic signal is seen even in the raw data. In order to 
emphasize the effect, they separated all genera into two categories:
short-lived (with the first and last occurrence dates being separated by
45~My or less) and long-lived. Only short genera shows the periodic
variation, although with less statistical significance than the
de-trended data. The raw short-lived genera data overlaid with the
CR flux is shown in Figure \ref{fig:s1}. 
The cross-correlation coefficients and the statistical 
significances for both data sets are given in Table \ref{t:1}.
Clearly, both data show correlation at very high statistical 
significance.

Spectral analysis by Fourier Transform builds on the result 
that almost any mathematical function can be decomposed into 
a sum of sinusoids.  The RM cyclicity result does not imply 
that the diversity record is sinusoidal, but that it does contain 
one or more components around 62 My in period which are anomalously 
large.  Our model would explain the basis of this large component; 
our cross-correlation result implies that the model can explain 
about half the overall variance in the fossil record.

Our model would predict the long-term variation of diversity, with a period
of about 64~My. The data, however, contains all time-scales.
The Fourier harmonic spectrum
shown in RM contains, in addition to the two cycles, a long tail
of short-frequency harmonics. This short-frequency component is largely
dominated by data binning, which sizes vary from 1~My to 9~My, and 
contaminate the data set. Therefore, we performed a separate 
statistical analysis of the data, in which short-time variations
are filtered out. We applied three different filtering techniques
to the de-trended data from RM paper.
A low-pass filter performs a forward Fourier transform to
calculate the spectrum, sets all Fourier harmonics with frequencies greater
than 1/(35~My) to zero and then performs the inverse Fourier transform
to restore the signal. A narrow window filter uses the same technique, but 
now keeps harmonics only within a narrow window around the 62~My peak.
The weighted window filter is analogous to the narrow window one, but
now the diversity spectrum is weighted with (multiplied by) the normalized
spectrum of the Solar motion, $z(t)$, which does show a prominent
harmonic peak around 63~My. The correlation coefficients
and the statistical significance levels are given in Table \ref{t:1}.
Overall, data filtering increases the correlation substantially; the
statistical significance rises to nearly 100\%. All these results confirm 
that our CR model describes the long-term variation of diversity very well.  
Note that for the weighted window filter technique, the filtered
data contains certain information on $z(t)$ and hence on the CR flux.
These two data sets --- the CR flux and the filtered data --- are not
statistically independent, therefore we do not evaluate the statistical 
significance of the correlation.  

It is also interesting to cross-correlate our CR model with the 
origination and extinction data sets separately. RM's
Fourier analysis of these sets shows that neither yields as strong
a 62~My-signal as the combined diversity data. They argued, therefore,
that the cycle is likely due to a combination of effects, rather
than just the extinction or just the diversification alone. 
Our study confirms this conclusion. The correlations of CR flux
with the origination intensity and with the extinction intensity from
RM data are very weak and statistically insignificant (p-value
is greater than 0.2 in both cases). These results are also summarized 
in Table \ref{t:1}. One must note, however, that if either had a phase offset
with respect to biodiversity, cross-correlation would be weak.  
\citet{Lieberman+07} have investigated this further.

\section{Correlation of diversity drops and CR maxima}
\label{a:corr2}

\begin{figure}
\epsfig{file=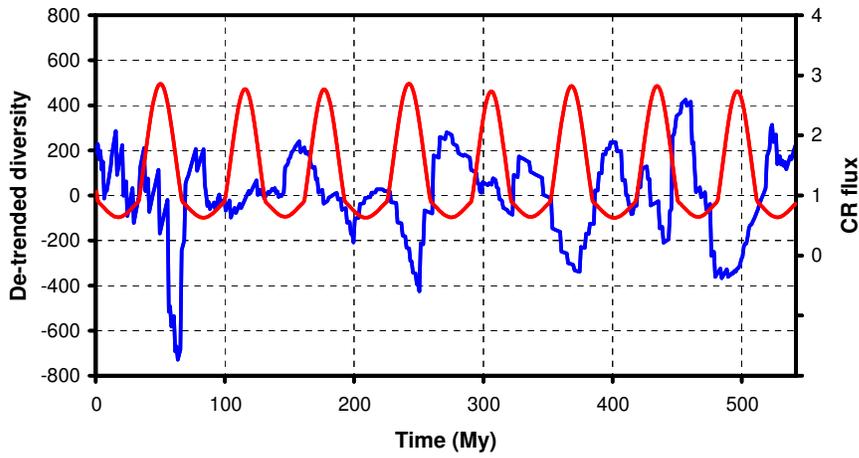, width=122mm}
\caption{The diversity variation (from \citealp{1}) and extragalactic 
cosmic ray flux at the Earth calculated from our model.
The de-trended diversity variation ({\it blue curve, left scale})
as a function of time over-plotted with the normalized cosmic ray flux 
calculated from our model ({\it red curve, right scale}). There are no
fit (and free) parameters in the model. The maxima in the cosmic 
ray flux coincide with minima of the diversity cycle.
Note also that the onset times of the diversity decline coincide with
moments of the rapid increase of the flux.}
\label{fig:3}
\end{figure}

Our model predicts that a higher CR flux should result in a larger 
diversity drop. To check whether this prediction 
is confirmed by the available data, we used the following algorithm.
First, one finds all local maxima of the CR flux through
the entire domain of 542~My. The times at which the CR flux is at 
maximum in each cycle and the corresponding value of the flux amplitude
of variation, calculated as the difference of the values at 
maximum and the preceeding minimum, are 
given in columns two and three in Table \ref{t:2}. Application of the
same algorithm to the de-trended data yields inaccurate results, because 
subtraction of the cubic fit introduces a large number of spurious 
local minima (the whole curve becomes saw-tooth-like, as is seen from 
Figures in RM and our Figure \ref{fig:3}). A much more accurate 
way to find local extrema is to use the short-lived data instead.
Since the cubic fit describes the global trend on the time-scale 
of 500~My, its subtraction hardly affects the local structure and
we can use these $T_{min}$ and $T_{max}$ for further analysis of
both de-trended and short-lived genera sets. Thus, in the next step,
for each CR $T_{max}$ one finds the nearest local minimum 
and the nearest preceding local maximum in the diversity curve.
These values are given in columns four and five in Table \ref{t:2}.
For all extrema, except just one, the data has the resolution coarser
than 1~My (often 3-5~My). For such large data bins, one takes
the median value of $T$ for the bin.
It is interesting that {\it each} CR peak has a so-defined diversity drop
within the cycle (not a single cycle is missed). Moreover,
the CR maxima and diversity minima nearly coincide, within few My.
The CR maxima times and the diversity minima times are shown 
in Figure \ref{fig:s2} versus the cycle number. The correspondence
of the minima/maxima is remarkable. 

\begin{figure}
\epsfig{file=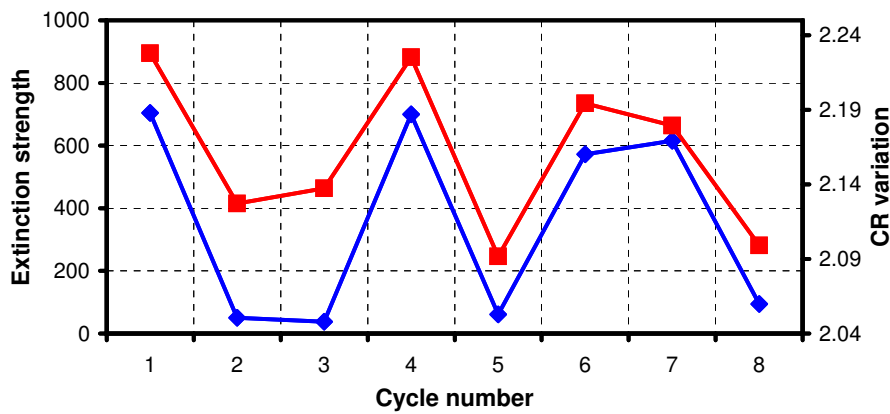, width=122mm}
\caption{The correlation of the extinction strength 
and the amplitude of the extragalactic cosmic ray flux.
The extinction strength ({\it blue curve, left scale}), 
calculated in each cyclic ``event'' in Figure \ref{fig:3} as the 
diversity drop from the preceding local maximum to the local minimum, 
and the amplitude of the EGCR intensity variation from our model
({\it red curve, right scale}) are plotted versus the cycle number 
(numbering is backward from the present).  
The correlation is about 93\% and has a probability less 
than one part in a thousand of arising from chance.}
\label{fig:4}
\end{figure}

\begin{figure}
\epsfig{file=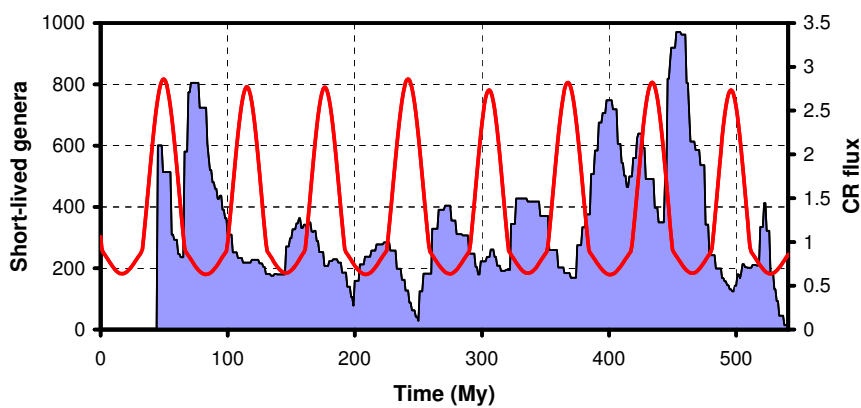, width=122mm}
\caption{The short-lived genera variation and extragalactic 
cosmic ray flux at the Earth calculated from our model.
The short-lived genera ({\it blue curve, left scale})
as a function of time over-plotted with the normalized cosmic ray flux 
calculated from our model ({\it red curve, right scale}). 
Note that the most of the minima in genera number nearly 
coincide with cosmic ray maxima, as in Figure \ref{fig:3}.}
\label{fig:s1}
\end{figure}

\begin{figure}
\epsfig{file=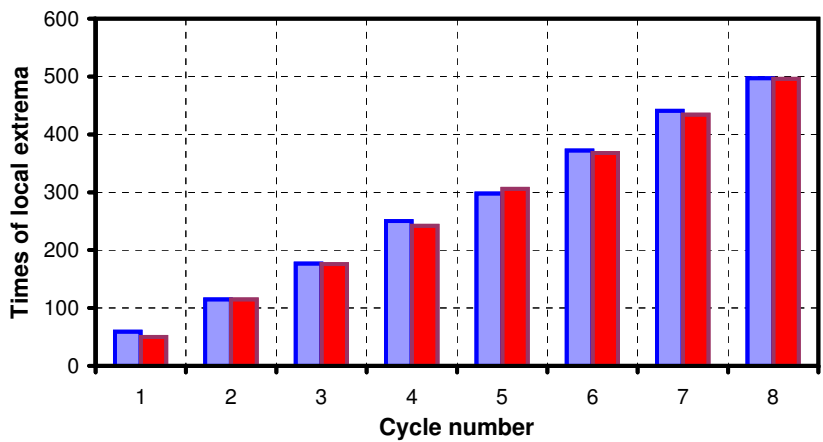, width=122mm}
\caption{ Diversity minima coincide with CR maxima.
The bar-chart shows the times of CR maxima, second column in 
Table \ref{t:2}, ({\it red bars})
and the times of the nearest local minima of diversity,
fourth column in Table \ref{t:2},
({\it blue bars}) versus the cycle number. 
The times of diversity minima are uncertain to few~My
due to large sizes of data bins (about 5~My, typically). 
\label{fig:s2}}
\end{figure}

\begin{figure}
\epsfig{file=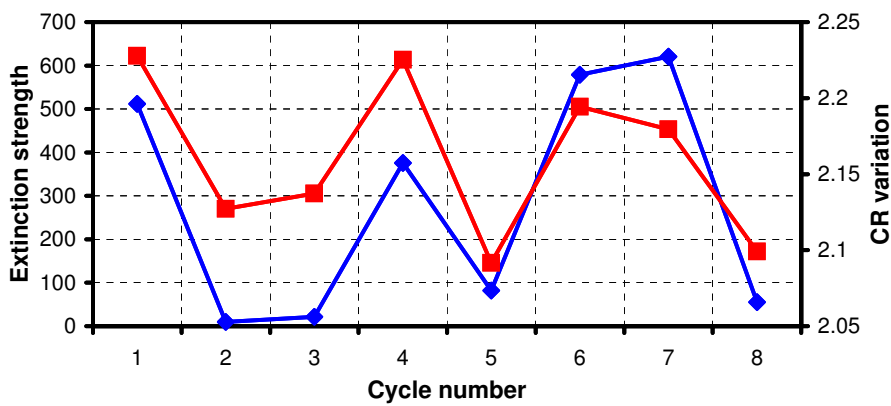, width=122mm}
\caption{The extinction strength of short-lived genera 
and extragalactic cosmic ray peak strength versus the cycle number.
As in Figure \ref{fig:3},
the extinction strength ({\it blue curve, left scale}), 
calculated as the diversity drop from each cycle
from preceding peak to minimum, and the relative EGCR  
intensity at maximum from our model
({\it red curve, right scale}) are plotted versus the cycle number 
(numbering is backward from the present).  
\label{fig:s3}}
\end{figure}

The drop in diversity, which we refer to as ``extinction strength'',
is defined as the difference in genera diversity at the maximum and the 
minimum, that is at times $T_{max}$ and $T_{min}$ given in columns 
4 and 5 of Table \ref{t:2}. The extinction strengths are calculated 
for both data sets, i.e., for the de-trended genera and the 
short-lived genera. They are given in the last two columns of 
Table \ref{t:2}. They are also plotted in Figures \ref{fig:4} and
\ref{fig:s3}. The correlation in both cases is very strong.
Although the short-lived genera set is not the 
``main'' sample --- neither in RM paper, nor in the present study, ---
it is remarkable that both samples show such strong correlations.
Thus, it justifies that the found correlations are real.
The correlation analysis shows, in particular, that CR maxima
are correlated with diversity drops with $r=0.80$ and $p=0.017$
(that is, 80\% correlation at 98.3\% confidence level) for 
the short-lived genera set, and with $r=0.93$ and $p=0.0007$
(that is, 93\% correlation at 99.93\% level, meaning that there is 
less than 0.07\% probability of the data-points happened to become 
``aligned'' this way by chance) for the de-trended data.

\clearpage

\begin{table}[ht]
\caption{Correlation data of the cosmic ray flux and 
various diversity data sets. The first group of data set
are data from \citet{1}. The second group represents de-trended
data filtered with the low-pass filter (to isolate long-term variations),
the narrow window and the $z(t)$-weighted window function (to isolate 
harmonics close to the 62~My cycle). In the latter case, 
the statistical significance and the p-value are not evaluated 
because the data sets (filtered and CR) are not independent.
The third group includes the origination and extinction data
from the supplementary information of \citet{1}.
\label{t:1}}
\begin{tabular}{p{1.5in}p{1.7in}p{1.4in}p{1.in}}
\hline
\hline
data set & cross-correl. (r-value) & stat. significance  & p-value\\
\hline
\hline
de-trended data & -- 0.49 & $\sim$ 100\% & $1.9\times10^{-7}$\\
short-lived genera & -- 0.32 & $\sim$ 99.9\% & $1.1\times10^{-3}$\\
long-lived genera & 0.11 & $\sim$ 70\% & 0.29\\
\hline
low-pass filter & -- 0.57 & $\sim$100\% & $6.4\times10^{-10}$\\
narrow window & -- 0.72 & $\sim$100\% & $2.0\times10^{-17}$\\
weighted window & -- 0.74 & --- & --- \\
\hline
origination intens. & -- 0.0039 & 3\% & 0.97 \\ 
extinction intens. & -- 0.067 & 44\% & 0.56\\
\hline
\hline
\end{tabular}
\end{table}
\begin{table}
\caption{Diversity drops and CR maxima. The columns are: 
(1) cycle number, (2) time at which CR flux is at maximum, 
(3) maximum value of the CR flux, (4) time of the local diversity 
minimum closest to CR~$T_{max}$,  (5) time of the closest preceding
local maximum of diversity, (6) extinction strength calculated 
from the de-trended data, (7) extinction strength calculated from
the short-lived genera data.
\label{t:2}}
\begin{tabular}{p{.5in}p{.7in}p{.7in}p{.8in}p{.8in}p{.7in}p{0.8in}}
\hline
\hline
Cycle ~~~~~~\# & CR: ~~~~~~$T_{max}$ & CR-flux ~~~~~~~~~~~ variation & Divers.: $T_{min}$  & Divers.: $T_{max}$ & Extinc.: ~~~~~~~~de-trend. & Extinc.: short-lived\\
\hline
\hline
1 &  50 My & 2.23 &  59 My &  74 My & 704 & 512\\
2 & 115 My & 2.13 & 115 My & 121 My & 50  & 10\\
3 & 176 My & 2.14 & 177 My & 184 My & 37  & 22\\
4 & 242 My & 2.26 & 250 My & 273 My & 700 & 376\\
5 & 306 My & 2.09 & 298 My & 308 My & 61  & 82\\
6 & 368 My & 2.19 & 372 My & 400 My & 572 & 579\\
7 & 434 My & 2.18 & 441 My & 454 My & 616 & 621\\
8 & 496 My & 2.10 & 497 My & 501 My & 94  & 55\\
\hline
\hline
\end{tabular}
\end{table}

\clearpage


\end{document}